\documentclass[10pt, conference, letterpaper]{IEEEtran}
\def\BibTeX{{\rm B\kern-.05em{\sc i\kern-.025em b}\kern-.08emT\kern-.1667em\lower.7ex\hbox{E}\kern-.125emX}}

\usepackage{etoolbox}
\DeclareListParser{\doslashlist}{/}
\newcounter{ndnNameComponentCounter}%
\newcommand{\ndnName}[1]{{%
  \setcounter{ndnNameComponentCounter}{0}%
  \renewcommand{\do}[1]{{%
    \ifnumgreater{\value{ndnNameComponentCounter}}{0}{\allowbreak}{}%
    \ifnumodd{\value{ndnNameComponentCounter}}{}{}%
    /##1}%
    \stepcounter{ndnNameComponentCounter}}%
``{\fontfamily{Arial}\selectfont\doslashlist{#1}}''%
}}

\usepackage{nicefrac}
\usepackage{siunitx}
\usepackage{array,framed}
\usepackage{booktabs}
\usepackage{todonotes}
\usepackage{gensymb}
\usepackage{
  color,
  float,
  epsfig,
  wrapfig,
  graphics,
  graphicx,
  subcaption
}
\definecolor{auburn}{rgb}{0.43, 0.21, 0.1}
\definecolor{arylideyellow}{rgb}{0.91, 0.84, 0.42}
\definecolor{applegreen}{rgb}{0.55, 0.71, 0.0}

\usepackage[normalem]{ulem} 
\usepackage{xspace}
\usepackage{textcomp,amssymb}
\usepackage{setspace}
\usepackage{comment}
\usepackage{latexsym,fancyhdr,url}
\usepackage{enumerate}
\usepackage{algorithm2e}
\usepackage{algpseudocode}
\usepackage{graphics}
\usepackage{xparse} 
\usepackage{xspace}
\usepackage{multirow}
\usepackage{csvsimple}
\usepackage{balance}
\usepackage{color,soul}

\usepackage{amssymb}

\usepackage{
  tikz,
  pgfplots,
  pgfplotstable
}

\usetikzlibrary{
  shapes.geometric,
  arrows,
  external,
  pgfplots.groupplots,
  matrix
}

\pgfplotsset{compat=1.9}

\usepackage{mathtools,}

\DeclareMathAlphabet{\mathcal}{OMS}{cmsy}{m}{n}

\DeclareGraphicsExtensions{%
    .png,.PNG,%
    .pdf,.PDF,%
    .jpg,.mps,.jpeg,.jbig2,.jb2,.JPG,.JPEG,.JBIG2,.JB2}

\begin{document}
\fancyhead{}
\def\thetitle{An Information Centric Framework for Weather Sensing Data}
\title{\thetitle}

\makeatletter
\def\ps@IEEEtitlepagestyle{%
  \def\@oddfoot{\mycopyrightnotice}%
  \def\@evenfoot{}%
}
\def\mycopyrightnotice{%
  {\footnotesize \textcolor{red}{\begin{tabular}[t]{@{}l@{}} This paper has been accepted for publication by a workshop held in conjunction with the IEEE International Conference on Communications. © 2022 IEEE. \\ Personal use of this material is permitted. Permission from IEEE must be obtained for all other uses, in any current or future media, including reprinting/republishing \\ this material for advertising or promotional purposes, creating new collective works, for resale or redistribution to servers or lists, or reuse of any copyrighted \\ component of this work in other works.\end{tabular}}}
  \gdef\mycopyrightnotice{}
}

\author{
\IEEEauthorblockN{
Robert Thompson\IEEEauthorrefmark{1},
Eric Lyons\IEEEauthorrefmark{2},
Ishita Dasgupta\IEEEauthorrefmark{2},
Spyridon Mastorakis\IEEEauthorrefmark{3},\\
Michael Zink\IEEEauthorrefmark{2},
Susmit Shannigrahi\IEEEauthorrefmark{1}}
\IEEEauthorblockA{\IEEEauthorrefmark{1}Computer Science Department, Tennessee Tech University, Cookeville, TN}
\IEEEauthorblockA{\IEEEauthorrefmark{2} University of Massachusetts Amherst, USA}
\IEEEauthorblockA{\IEEEauthorrefmark{3}Computer Science Department, University of Nebraska at Omaha, USA}
}


\date{}

\maketitle
\begin{abstract}


Weather sensing and forecasting has become increasingly accurate in the last decade thanks to high-resolution radars, efficient computational algorithms, and high-performance computing facilities. Through a distributed and federated network of radars, scientists can make high-resolution observations of the weather conditions on a scale that benefits public safety, commerce, transportation, and other fields. While weather radars are critical infrastructure, they are often located in remote areas with poor network connectivity. Data retrieved from these radars are often delayed or lost, or even lack proper synchronization, resulting in sub-optimal weather prediction.
This work applies Named Data Networking (NDN) to a federation of weather sensing radars for efficient content addressing and retrieval. We identify weather data based on a hierarchical naming scheme that allows us to explicitly access desired files. We demonstrate that compared to the window-based mechanism in TCP/IP, an NDN based mechanism improves data quality, reduces uncertainty, and enhances weather prediction. 
Our evaluation demonstrates that this naming scheme enables effective data retrieval, while compared to the window-based mechanism in TCP/IP, an NDN based mechanism improves data quality, reduces uncertainty, and enhances weather prediction. 

\end{abstract}


\section{Introduction}
\label{sec:intro}



Weather sensing and forecasting has become increasingly accurate in the last decade thanks to high-resolution radars, efficient computational algorithms,  and  high-performance computing facilities. The NSF Engineering Research Center for Collaborative Adaptive Sensing of the Atmosphere (CASA) studies the lower atmosphere with networks of high resolution Doppler weather radars with the goal to improve severe weather awareness \cite{MPC+09}. Currently, CASA operates a network of seven X-band radars in the Dallas/Fort Worth (DFW) Metroplex~\cite{casadfw}. These radars have mechanically steered antennas and are tasked to perform surveillance scans (antenna rotates a full 360$^\circ$) at different elevation angles. 

These radars are currently connected to a data collection facility at National Oceanic and Atmospheric Administration (NOAA) in the DFW area. Being located in remote areas, the radars face significant network connectivity challenges. First, the available bandwidth varies considerably across these radars. Some radars are connected to high-speed networks with hundreds of Mbps connectivity, some share a 10Mbps connectivity with competing traffic, and some have dedicated, guaranteed 10Mbps connections. The other problem is congestion on the links - as the weather becomes worse, these radars produce more data and there is increased user demand. Given the limited amount of available bandwidth, uncoordinated requests (e.g., requesting all available files from a radar at once) create congestion and delays data retrieval from the radars which, in turn, delays short term weather forecasts. Finally, data from all radars are combined into a mosaicked product before being handed off to the weather prediction workflows. Due to the difference in data generation rates of the radars (radars from different vendors have different rotational speeds), different network speeds, and network congestion at a given moment, the files from these radars arrive at the merging site (NOAA) at different points in time. Merging files that are not temporally synchronized may reduce the quality of the merged product and affect weather prediction.

In this work, we move away from TCP/IP's push-based model that utilizes a static distribution model in favor of NDN's pull-based model. Once the files are generated at the radar site, they are currently pushed to NOAA's computing facility for processing. In this work, we utilize Named Data Networking (NDN)~\cite{Zhang:NDN:CCR2014} that allows the data collection facility to request the necessary files from the radar for processing. 
To enable efficient, NDN-based communication, we develop and utilize a hierarchical naming scheme that explicitly captures radar parameters. 
Rather than downloading everything that is available from the radars, this naming scheme allows the merging site to specify which exact datasets it requires. To facilitate this, we develop a signaling mechanism where we piggyback information about new files as they are generated.


Finally, we utilize ndnSIM~\cite{mastorakis2017ndnsim} to create an exact topology of the CASA radar federation, complete with accurate delay and bandwidth information. We also utilize actual files generated by these radar sites for our simulation. Our simulation shows that an NDN based system improves the quality of data, reduces congestion on bottleneck links, and improves weather prediction workflows. Indeed, we demonstrate that utilizing an NDN based method to retrieve data uncovers a hazardous weather event (a tornado) that would be harder to detect with the current dataflow model.

Our work is novel in several ways. 
We develop a name-based protocol that allows us to retrieve weather data deterministically over networks with different connectivity. By naming files to represent the actual rotation of the radars, we allow the consumer to request the exact files it needs for weather prediction, rather than periodically downloading an arbitrary number of files. If necessary, our protocol allows us to explicitly notify the consumer when a new file is available. 
By integrating an NDN-based naming and data retrieval strategy, we demonstrate that NDN's pull based model enables more accurate weather predictions compared to TCP/IP's push based model.


\section{Background}
\label{sec:background}

\subsection{Why use NDN over HTTP or TCP/IP based protocol?}
In federated networks of radars, the radars may communicate with clients through paths consisting of several network hops. These paths may span different service providers, become congested over time, have certain bandwidth constraints, while the overall connectivity between radars and clients may be intermittent. TCP/IP relies on end-to-end connections, without any support from the network infrastructure. 


On the contrary, NDN operates on names, data can be delivered from anywhere, automatic in-network caching speeds up data delivery that reduces duplicate requests and speeds up retransmissions. Finally, when several routes are present, NDN's hop-by-hop forwarding can bypass failure and intelligently pick one or more routes based on observed traffic patterns.


NDN can also facilitate fast retransmissions through in-network caching. At the same time, solutions such as SSL/TLS which are widely used in TCP/IP, secure the communication channels, largely depends on the underlying connectivity. On the other hand, NDN makes security a property of the data itself, decoupling it from the underlying connectivity.

CASA was formed to study the lower atmosphere with networks of high resolution Doppler weather radars with the goal to improve severe weather awareness \cite{MPC+09}. The volumetric data produced by these continuously operating remote sensors must be distributed to processing servers quickly and efficiently, such that analysis can occur in near real time for the sake of warning the public of fast developing threats such as tornadoes and high winds. The networked radar concept requires that asynchronous raw data from multiple sources is blended together to create value-added meteorological products. At any given time the characteristics of the ongoing weather regime determine the necessity and priority of certain products. For example, a hail detection algorithm takes on high importance only when strong thunderstorms are ongoing, whereas forecasting algorithms may be of more importance well in advance of such severe weather events and perhaps somewhat less so once the event has started. Currently, CASA operates a network of seven X-band radars in the DFW Metroplex~\cite{casadfw}. These radars have mechanically steered antennas and are tasked to perform surveillance scans (antenna rotates a full 360$^\circ$) at different elevation angles. Since the radars are not identical, their form of producing atmospheric data is slightly different, and data generation is not synchronized across all radars (e.g., the execution of a surveillance scan does not take the same amount of time for all radars). 

In Section~\ref{subsubsec:dataflow}, we provide an overview of the current data flow paradigm and Sect.~\ref{subsubsec:asyncdata} outlines the challenges related to this approach and motivates how Information Centric Networking can mitigate these challenges.

\subsection{Current Dataflow Paradigm}
\label{subsubsec:dataflow}
The underlying data transport mechanisms used in the CASA network are based on the traditional TCP/IP protocol stack. For example, UCAR's Local Data Manager (LDM) system~\cite{unidataldm} is used for event based distribution and analysis of radar data. The event based distribution relies on a sender-driven pub/sub system, whereby data requests from downstream clients are registered with regex-like pattern matching schemes based on expected file naming conventions.  Data filenames generally include product or radar name, valid time, and the file format abbreviation as suffix.  Client side data requests include the IP address or DNS name of the upstream server associated with each product pattern and these are contained in a configuration file.  The upstream data server must include a corresponding configuration entry allowing the downstream client IP address/DNS name to request such data.  Any changes to the configuration files on either client or server side require a program restart to take effect, during which all data ceases to flow for several seconds. Thereafter, any data arriving to or input from the server side matching the client's requested pattern is forwarded to the client.  Data flows and the applications using the data are disjointed in this respect.  This sender-driven approach is not well suited for mostly data-driven algorithms since a priori knowledge of active algorithms and their data needs is required, and modifications to the data retrieval service is disruptive to the overall system and potentially other users. 
The paradigm of preconfigured data retrieval on a per machine basis is not well suited for the virtualized, cloud-based, highly adaptive compute resources that are used in the CASA system today~\cite{escience2019}. An approach that eliminates the need for a priory knowledge of the applications compute resource and data requirements will benefit current and future CASA data processing workflows.

\subsection{Weather Sensing}
\label{subsec:rwcasa}
In contrast to the nationwide NEXRAD radar network operated by the National Weather Service (NWS) which consists of 160 homogeneous radars, the CASA system in DFW consists of a heterogeneous and federated network of radars. While in the NEXRAD system there is only little atmospheric volume that is covered by more than one radar, a significant portion of the coverage area of the CASA system is covered by 2 (or in some cases 3) radars. Thus, merging the individual radar data in a mosaic in a timely manner is important for algorithms that use these merged data for weather product generation.

Current radar networks (including NEXRAD) make use of data distribution applications like LDM~\cite{unidataldm} that are implemented on top of the traditional TCP/IP stack. Contrary, the work presented in this paper presents a new approach based on NDN to make radar data distribution and the execution of weather algorithms more efficient. Zhang et al.~\cite{Zhang:Infocom:2017} have presented a synchronization protocol that is also designed for the distribution of weather data. While the work presented by Zhang et al. focuses on the distribution of weather products to the end users, our approach focuses on the generation of such products by providing a new approach for transmitting data from the radars to a central compute site where weather product algorithms are running. In addition, our design and proposed naming scheme for data generation and retrieval in rounds is inspired by RoundSync, a protocol for distributed dataset synchronization in NDN~\cite{de2017design}.

\subsubsection{Asynchronous Data}
\label{subsubsec:asyncdata}
The operation of the radars in the CASA network is not synchronized in any respect, however the data they generate still needs to be mosaicked at a central processing location, and there is still a notion of an ideal set of data files that should be ingested for each mosaic.  In general, a mosaic is a full volumetric set of data from each radar, as closely linked in time as possible.  Radars produce volumes made up of multiple data files at intervals from 50 to 70 seconds.  It is less than ideal to not include some of the data files making up a volume in the mosaicking process, but also less than ideal to include data files representing the same portion of the volume at multiple times, which results in a smearing effect. Time based windowing, whereby an algorithm waits for a specified period of time for data from all radars to arrive, then executes based on the input it received quite often leads to one or the other of these results. Moreover a result we always strive to avoid is to not include any data from a radar in a mosaic, starving regions of the grid of all data and possibly causing users to make ill informed decisions.  Therefore the time windows must be kept open longer than the ideal interval, so as to guarantee in all normal  network traffic situations that at least one file from each radar shall arrive and be included in the mosaic.  A better solution would be to request all data files making up a volume from each radar in an explicit fashion. Current time-based naming schemes can make this problematic to request, so we propose an alternative scheme indicative of the periodic volumetric collections of data files from each radar and explain it in detail in Section~\ref{subsec:scheme}.

\section{Methodology/Design}
\label{sec:methodology}


\begin{figure}
  \centering
  \includegraphics[width=0.7\columnwidth]{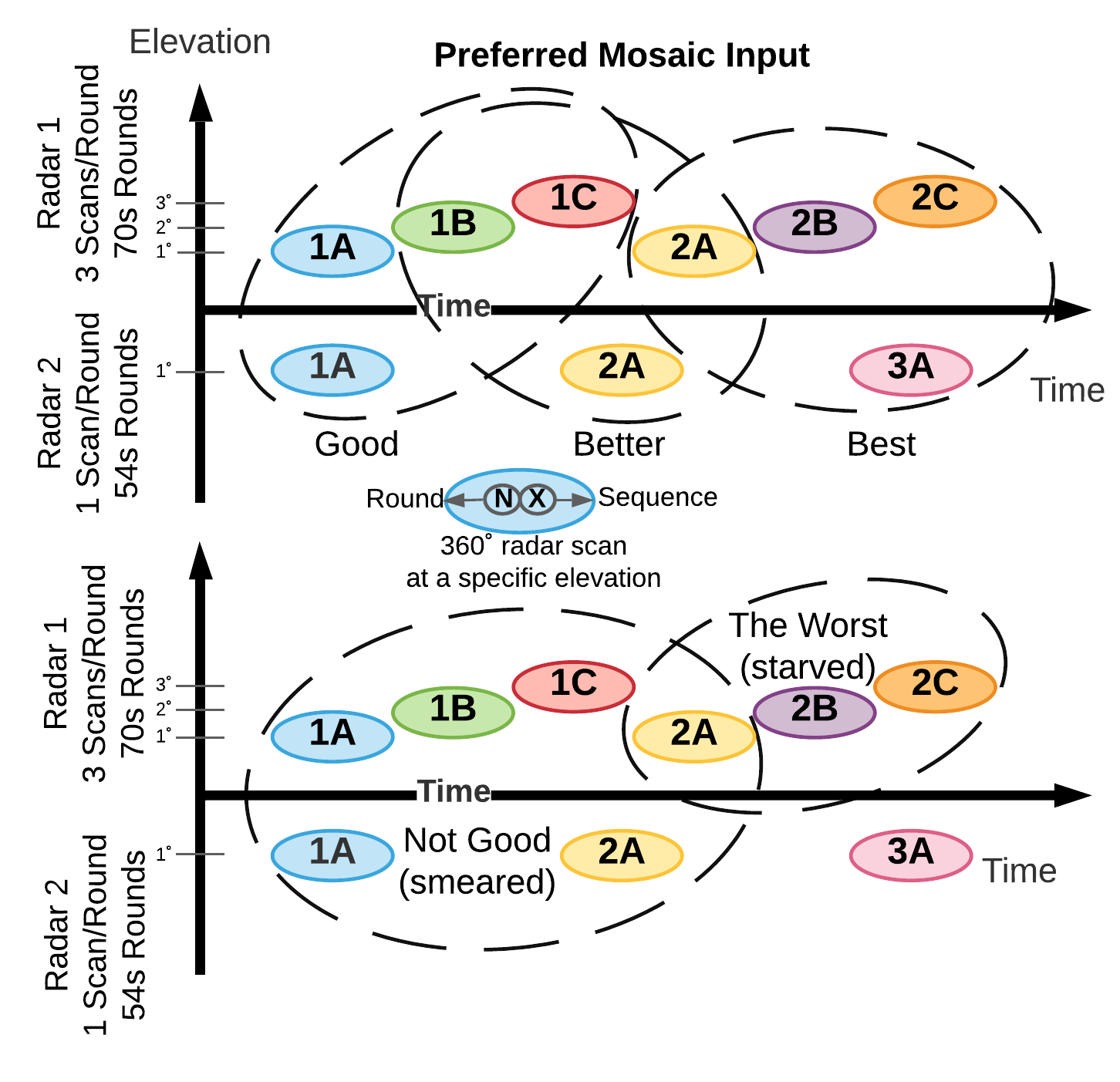}
  \caption{Illustration of the input for the merging algorithm to create a mosaic from data generated by two heterogeneous radars. Top figure shows the preferred scenarios while the bottom figure shows worst-case scenarios.}
  \label{fig:merging}
\end{figure}

\subsection{Scenario}

Tha CASA system generates dozens of meteorological products in near real time. Some of these products are generated 24/7/365, others on demand, based on the characteristics of the ongoing weather regime.  First order processes include calculating rainfall rate and accumulations, short term nowcasts (0-30min), hydrometeor classifications (rain/hail/snow), hydrological products (runoff, streamflow), and network wind products.
In addition, various post-processing routines operate on the gridded product data, including raster image generation, contouring, format conversions, and end user driven, GIS based data extraction. Timely generation of these products is essential for the warning process and requires significant network and compute resources. 

For better illustration of the design of our information-centric approach, we focus on the process of mosaicking data from individual radars into a merged grid, as shown in Fig.~\ref{fig:merging}. We have chosen this example since it requires a minimum amount of data from each individual radar to generate an accurate grid. The accuracy of the gridded data impacts the performance of weather algorithms that perform on gridded data. Ideally, data from all radars would arrive simultaneously at the processing node to allow an immediate generation of the merged grid. However, as described before, radars and the access networks connecting them to the Internet are heterogeneous.  Additionally, competing background traffic means available bandwidth are different across these radars. This heterogeneity may result in suboptimal merged products - if a less than ideal number of files are included in the merged product, it may result in a worst case scenario (bottom part of Fig.~\ref{fig:merging}). When too many files are included in the merged process it may end up ``smearing" the merged product, again resulting in a suboptimal product.  A suboptimal merged product might impact weather prediction (e.g., decision of a weather forecaster issuing a tornado warning or not) as we will demonstrate in Sect. \ref{sec:eval}.




We present a new design for the pull-based retrieval of individual radar data that is based on NDN to allow for a more efficient transport, and better results for downstream meteorological products. 
We utilize rounds (a complete volume scan by the radar) and sequence numbers (files generated within a round). For example, a particular radar might make a complete 360$\degree$ rotation in a minute. It may also create 3 files for the whole round, they can be sequence 1,2, and 3, each containing approximately 20 seconds worth of data. 
The use of round and sequence numbers allows the radars to operate independently. For example, each radar may have their own rotational speed (RPM)  that does not need to be synchronized. Additionally, this scheme allows the radars to dynamically change the rotation speed or even allow them to go offline without the need for synchronization with the other radars.

\subsection{Data Retrieval in Rounds}\label{subsec:rounds}
Currently, the weather data collected from the radars are named as \ndnName{location-state.YYYYMMDD-HHMMSS.netcdf.gz}. One such file on the radar might be named as  \ndnName{addison.tx-20200909-000056.netcdf.gz}. This naming convention provides sufficient information for subsequent processing - information such as location and time allows the workflow to analyze files from different radar sites, merge all files over a given time window, and run subsequent computations on them. However, the radars have a fixed rotational frequency (e.g., three times every 70 seconds for radar 1 shown in Fig.~\ref{fig:merging}), which is not captured by the time-based file names. The generation of files is not connected to a particular round. 
When using time-based windows, a computational workflow might end up with truncated data (e.g, when a full round of data is not captured in the file) or miss data from a few rounds (e.g., the last rotation was not captured in the netcdf files with the time window).

Moving from implicit naming where the workflows will have to make assumptions or look into the actual data is cumbersome. In this work, we transition from implicit naming (time based) to explicit naming (round based) where each file generated by the radars are named as \ndnName{/data/radar1/\_round=1/\_seq=A} (first data file for radar 1 in the example shown in in Fig.~\ref{fig:merging}). 
For workflows that need timestamps, we can simply add the timestamp to the name - \ndnName{/addison/tx/radar1/\_round=15/\_seq=7/YYYYMMDD/HHMMSS}. 

This naming scheme allows the workflows to look at the names and choose the most relevant data. One example might be when a workflow looks at all rounds collected up to a certain time. The NDN naming and data retrieval makes it much easier to retrieve content based on actual radar rotation, rather than using timestamps. Further, these names allow the clients to predict the exact names of the future data (e.g. \ndnName{/data/radar1/\_round=1/\_seq=C}, simplifying the application and workflow logic.

\begin{figure}[!t]
    \centering
    \includegraphics[width=0.7\columnwidth]{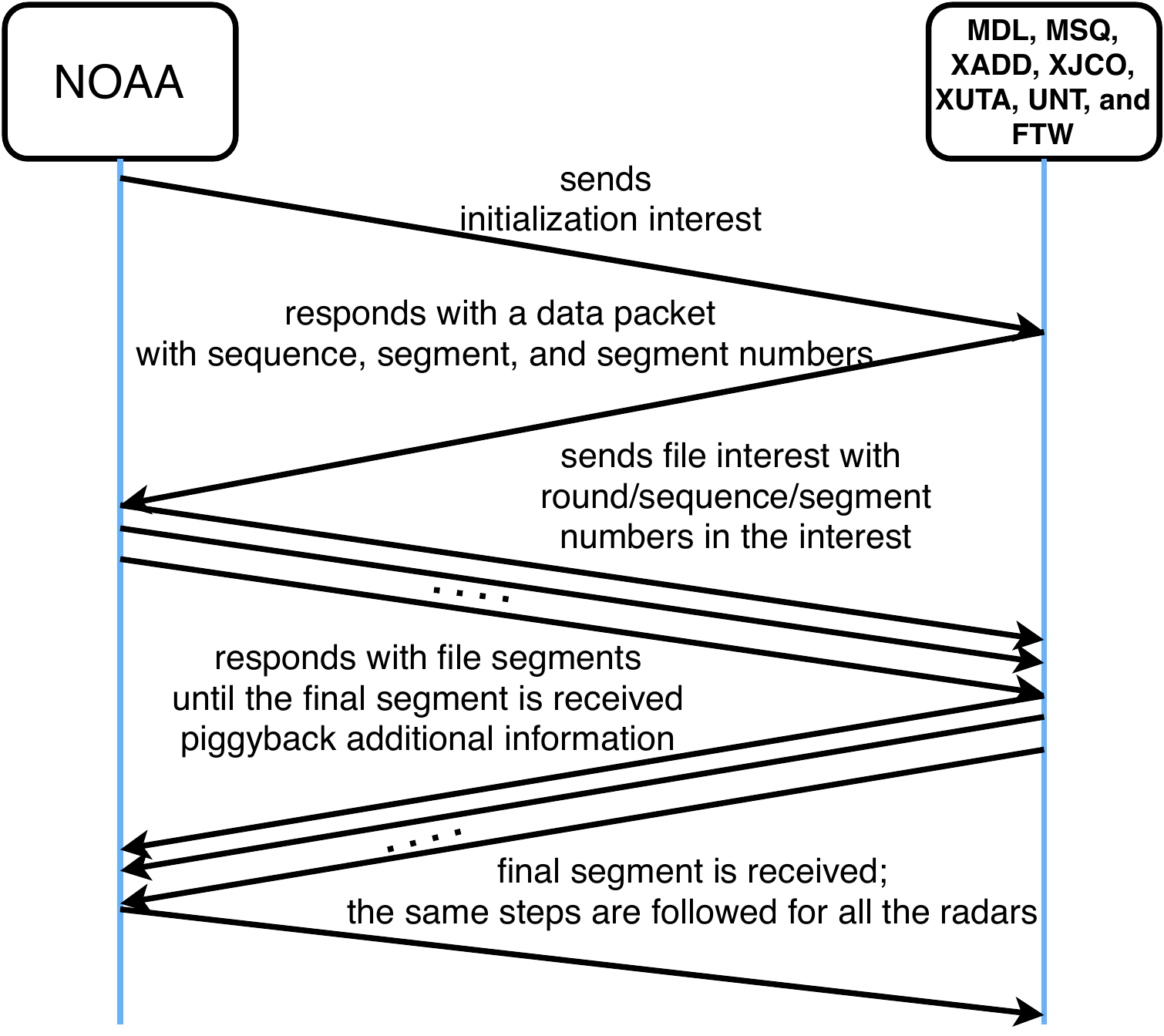}
    \caption{NDN based Interest/Data Exchange}
    \label{fig:protocol}
\end{figure}

\subsection{A Naming Scheme for Periodic Radar Data}
\label{subsec:scheme}
NDN clients indicate a request for data via an Interest packet. In this work, we utilize two types of Interest - initialization Interests and normal Interests. The clients send initialization Interests and query the radars about their current states. On receiving this Interests the radars return their current state of data collection, specifically the current round and the sequence number. The current round can be derived from current time (time since epoch modulo $n$) or based on some other numbering scheme. There are a fixed number of sequences in a round depending on the rotational speed of the radar (3 sequences for radar 1 and 1 sequence for radar 2 in the example shown in Fig.~\ref{fig:merging}).

Figure \ref{fig:protocol} shows the actual Interest/Data exchange protocol. The initialization Interests assumes the following format -\\ \ndnName{/Interest/radar\{1..n\}/\_current\_round/current\_seq/number\_of\_files}. On receiving this Interest, the radar returns the most recent file with the current round number and the current sequence number. The client then may start requesting the files by their actual names, e.g.,  \ndnName{/data/radar1/\_round=15/\_seq=7}. We assume that the radars do not have large storage and only store a small number of files.  As a file is downloaded, the radar has the opportunity to piggyback information on the returning data packets, as we discuss in the next section.

\subsection{Piggybacking}

To inform clients about the name of the next file that will be produced by a radar, we design a piggybacking mechanism, where the file name is piggybacked to clients through the data of the current file. Specifically, this name is added to the metadata field of NDN Data packets. Once the client receives this information, it can launch another thread to start the parallel retrieval of the new file, before the retrieval of the current file has finished. This piggybacking approach can also be used in cases where the frequency of the radars change dynamically (e.g., when weather changes occur), so that clients stay informed about such changes.

This approach allows clients to adapt to changes that may happen on the radar side (e.g., generation of a new file, frequency changes). It comes at the cost of slightly increased sizes of certain Data packets, so that the required information can be encoded and piggybacked from radars to clients.



\section{Evaluation}
\label{sec:eval}

\begin{figure}[!t]
    \centering
    \includegraphics[width=0.7\columnwidth]{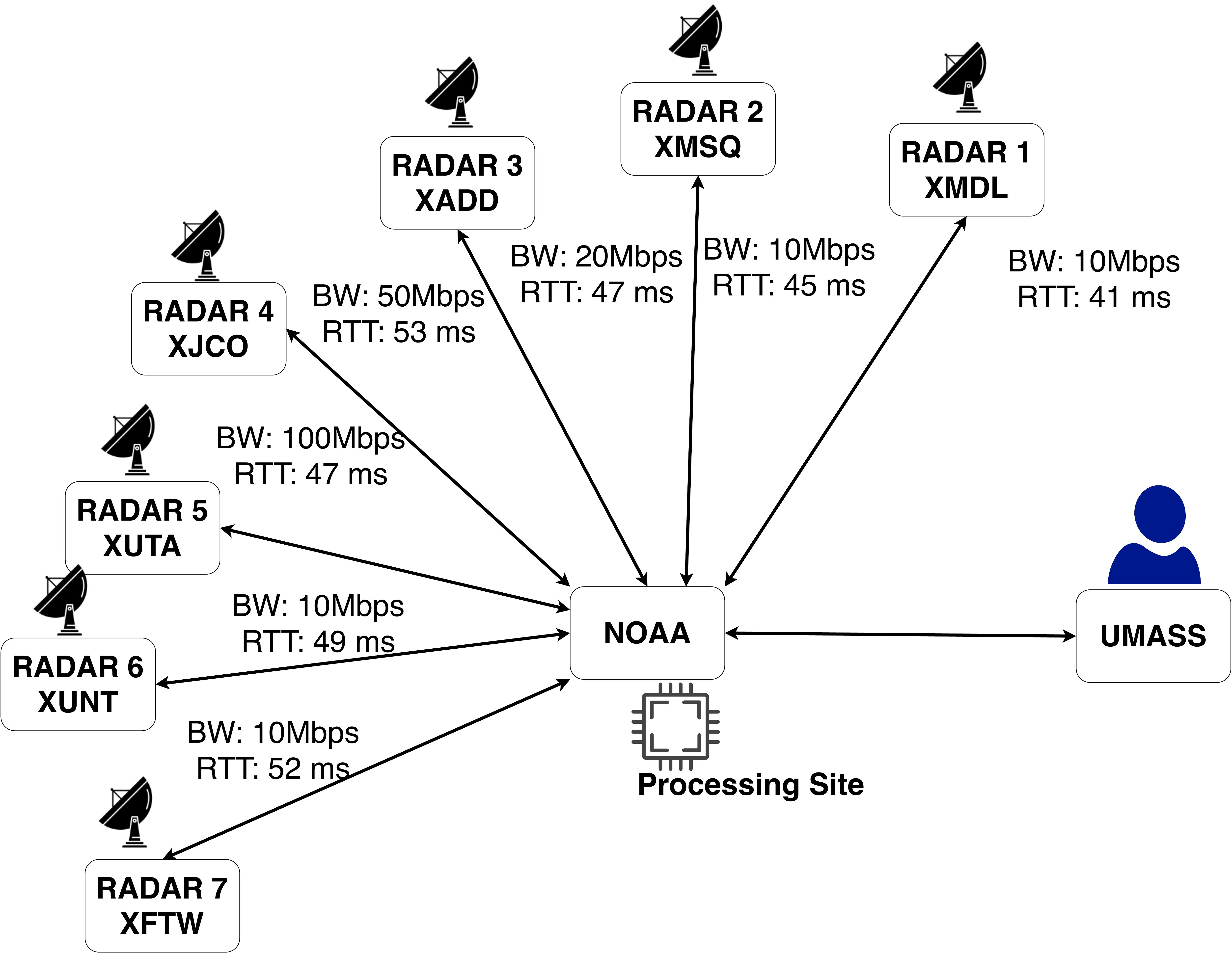}
    \caption{Experiment topology. The site names, bandwidth and delays are obtained from the actual CASA deployment.}
    \label{fig:simtopo}
\end{figure}


\begin{figure}[!ht]
\centering
 \includegraphics[width=0.8\columnwidth]{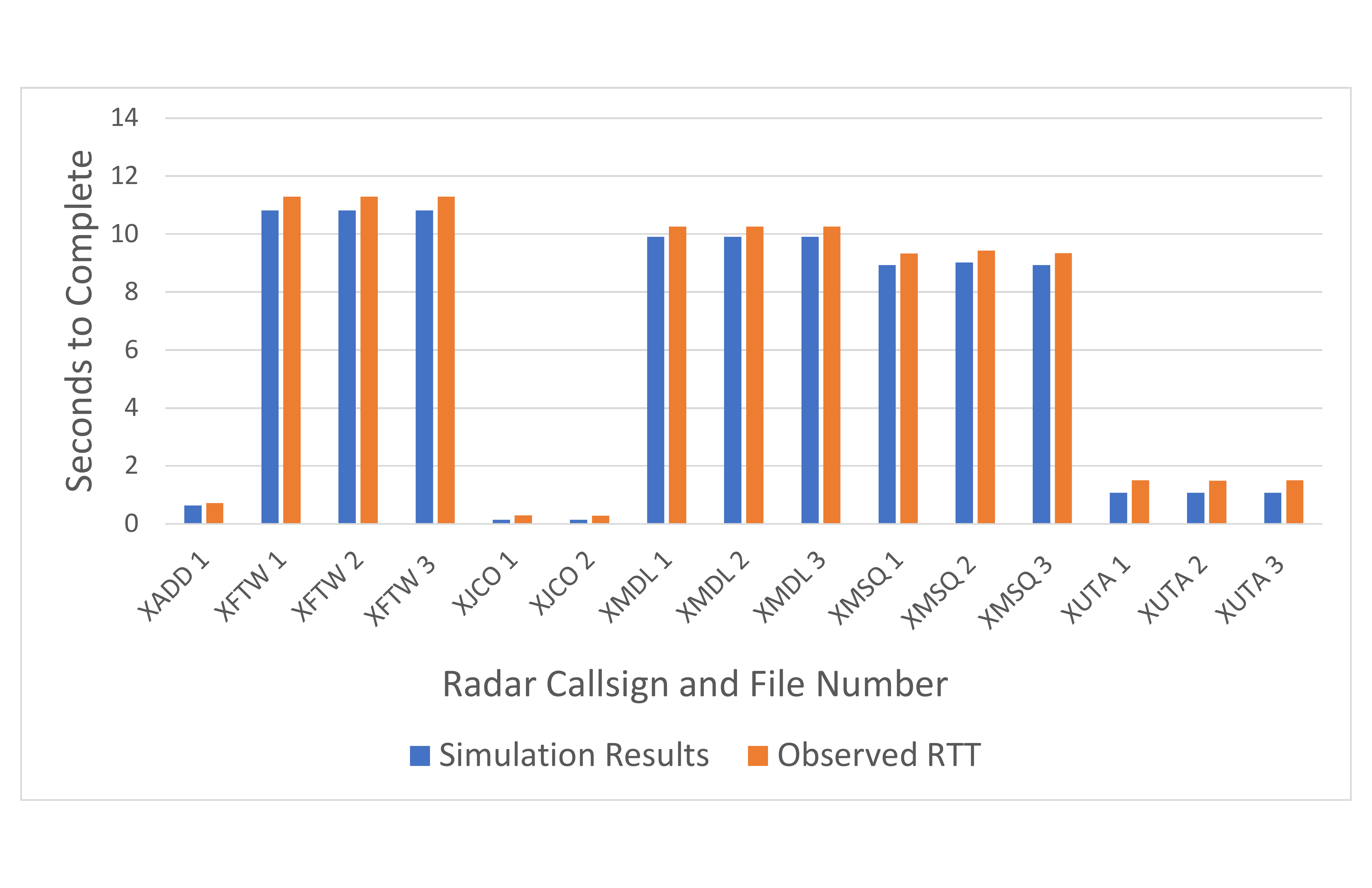}
   \caption{Time required for downloading datasets from radars using NDN. All datasets in this figure are needed for the weather prediction workflow. The observed RTTs from emulation are slightly higher than simulation. }
    \label{fig:sim_vs_emulation}
  \end{figure}

 \begin{figure}[!ht]
 \centering 
    \includegraphics[width=0.8\columnwidth]{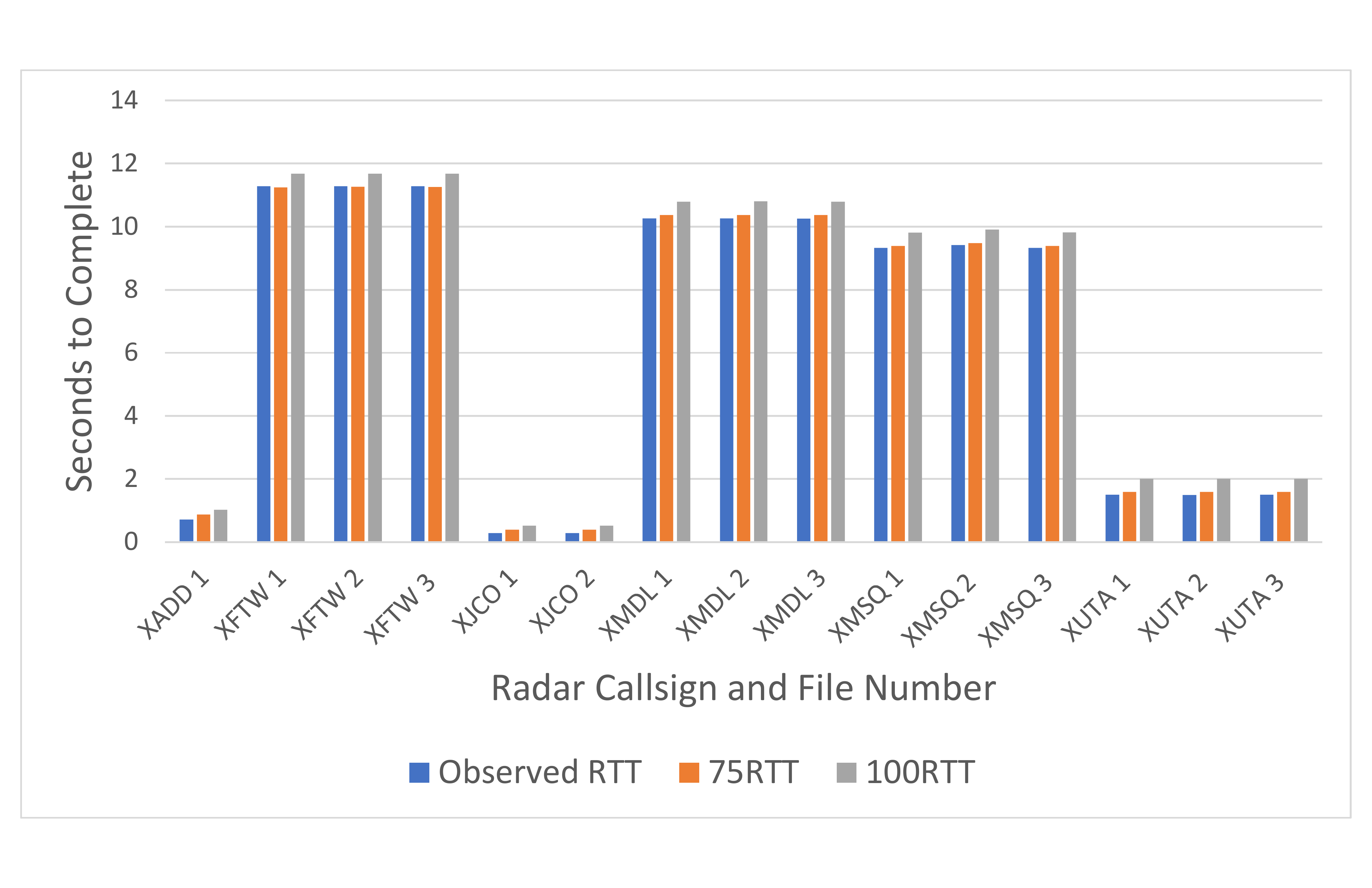}
    \caption{Effect of increased RTT on file download times }
    \label{fig:increasedRTT}
  \end{figure}

  \begin{figure}[!ht]
  \centering
    \includegraphics[width=0.8\columnwidth]{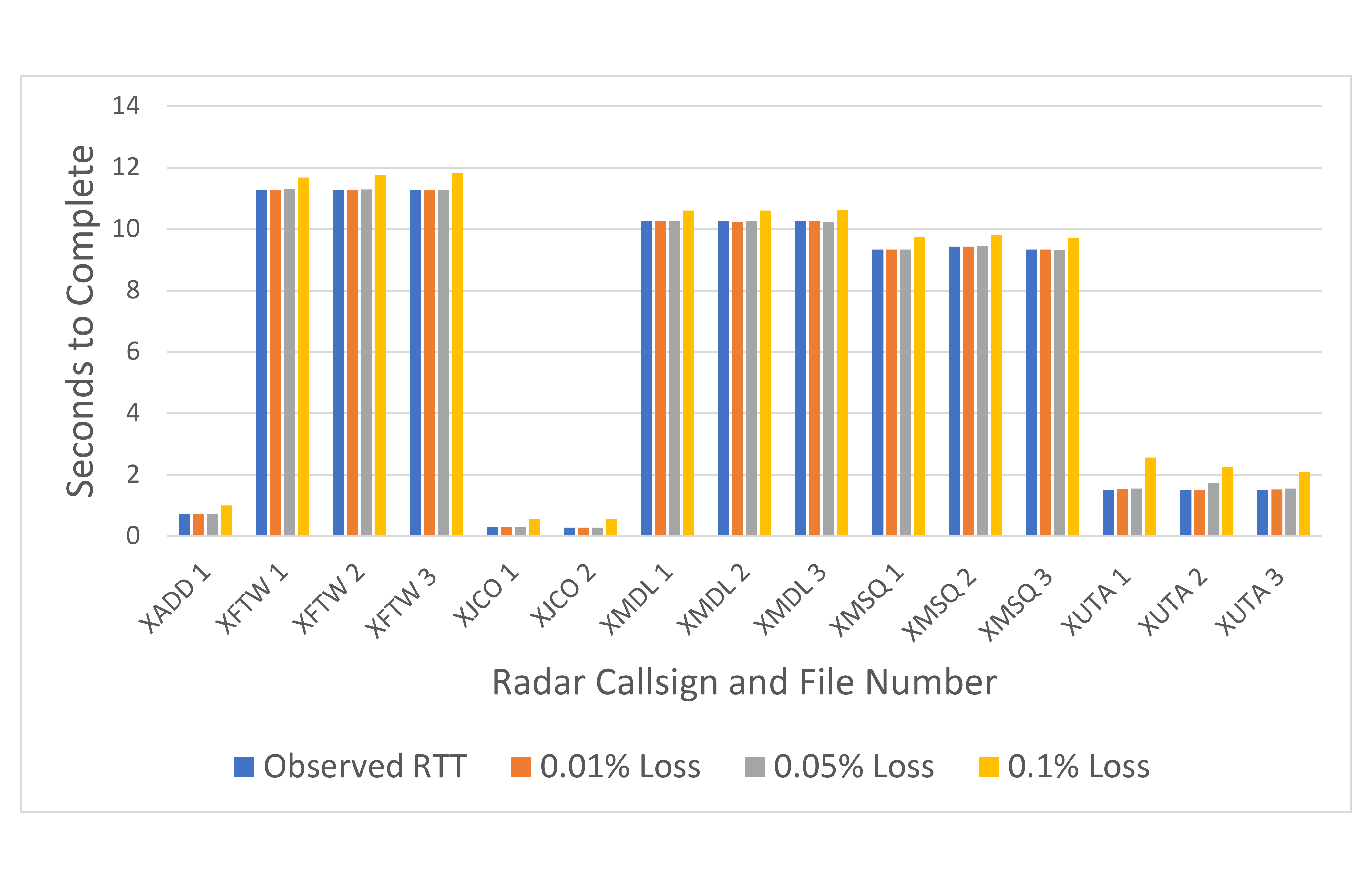}
    \caption{ Effect of increased loss on file download times}
    \label{fig:increasedLoss}
  \end{figure}

In this section, we present the evaluation of our information centric framework for delivering atmospheric data. In this work, we have utilized both simulation and a testbed of Virtual Machines (VMs) deployed on Google Cloud. We first describe the evaluation setup followed by a presentation of the results.


\subsection{Experiment Setup}


We used both simulation and emulation to evaluate our protocol. Figure~\ref{fig:simtopo} shows the topology we used for our experiment. We created this topology from the actual CASA radar topology. We use \textit{ping} between the NOAA and the radar sites to obtain the round trip time (RTT). The bandwidth numbers are obtained from operational experience. Note that some of these links are shared with other traffic (e.g., XMDL) while some links are dedicated (e.g., XUTA). We also looked at the actual request pattern obtained from CASA. We preserved the time between these requests and replayed the requests in real-time indicated by the actual logs. One radar was offline during our experiments so we did not use it for our experiments.

For simulations we utilizes ndnSIM~\cite{mastorakis2017ndnsim} and we used Virtual Machines (VMs) deployed on Google Cloud for emulation. One of the VMs was designated as the consumer with the other 6 as radars producing data. We used ndn-tools (ndncatchunks and ndnputchunks) to publish and retrieve data. We used a  script to determine when files were originally requested in the CASA log, and downloaded them using ndncatchunks.  For emulating bandwidth and latency on Google Cloud, we utilized the linux utility ``tc`` to set the delay and loss. During the scenarios with increased latency and loss, we changed the ``tc" parameters to accurately reflect the network condition. We tested each scenario 20 times for statistical significance.


As shown in Fig.~\ref{fig:simtopo}, NOAA is the processing site that communicates with the seven radars. It expresses Interest based on the naming scheme we described in Section \ref{sec:methodology}. 
The producer produced a file of a certain size at a certain time. We obtained the file sizes and the respective generation times from an actual data log at CASA. The consumer requests those file at times specified in the data access log.

\subsection{Results}
The baseline timings from each radar can be seen in Fig.~\ref{fig:sim_vs_emulation}. This baseline simulation consists of each radar starting the round at simulated second 0 and completing once all of the files from each radar are available on the consumer. The total transfer time begins once the file interest is sent out and finishes after the entire file is available on the consumer. The hosted file sizes are based on the file sizes from each radar collected during a weather event and the link bandwidth and latency match the ones shown in Fig. \ref{fig:simtopo}. While each link has a relatively low latency, the total transfer time is dependent on the hosted file size and link bandwidth. For example, the Cleburne file transfers take very little time due to the small file sizes (~.2 MB) and the link's high bandwidth (50Mbps). The opposite can be seen with the files generated by the Fort Worth radar, which are the largest of all radars (12MB) and the smallest available bandwidth (10Mbps). 

As the delay and loss increase the files take slightly longer to download.  The variance in time also goes up, especially with increased loss. However, the increase in download time is small since NDN enables fast retransmission from cache and does not need to decrease congestion windows as aggressively as TCP/IP. Figures \ref{fig:increasedRTT} and \ref{fig:increasedLoss} show the effect of increased loss and delay on transfer times. In this work we do not compare NDN with TCP/IP with increased loss and delay but refer the reader to our previous paper \cite{9335806}. In that work we show that NDN outperform TCP/IP even when a small amount of loss is present.





\begin{figure}[!ht]
    \centering
    \includegraphics[width=\columnwidth]{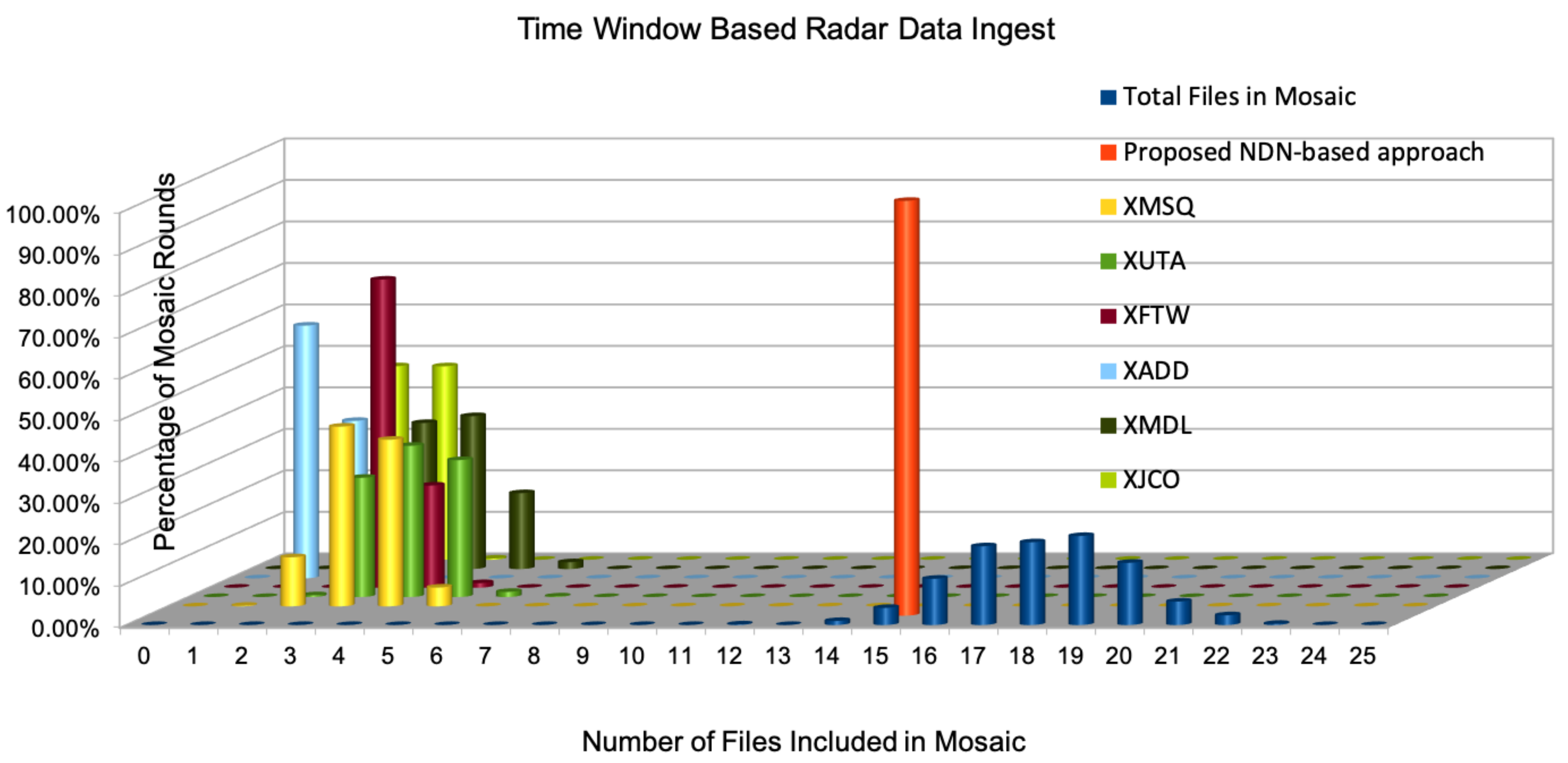}
    \caption{Current method vs NDN based file retrieval. The left bars show the number of files from each individual radar that are included in mosaics on a percentage basis. The blue bars show the cumulative files used for a mosaic. The red bar shows NDN's improvement, always the ideal number of files. One radar was offline and was excluded from the simulation.}
    \label{fig:time_window}
\end{figure}

Figure \ref{fig:time_window} shows the comparison between the current, TCP/IP based workflow and the proposed NDN based workflow. With the current time based window, a different number of files can be included in the mosaic since the client does not always receive all the required files from the radars, an artifact of TCP/IP's push based model. The left side of the figure shows only 40-50\% of the files are included in the mosaic due to the non-determinism of the time window based process. As a result, the total number of files included in a weather prediction workflow varies considerably (12-24) as the blue bars show on the right. The ideal number of files for this operation is 15 (all files from a round across all radars, see Figure \ref{fig:sim_vs_emulation}). On the contrary, NDN's pull based model always provides the ideal number  of the files (i.e., 15 in this case) for inclusion in the workflow.

\begin{figure}[!ht]
    \centering
    \includegraphics[width=0.7\columnwidth]{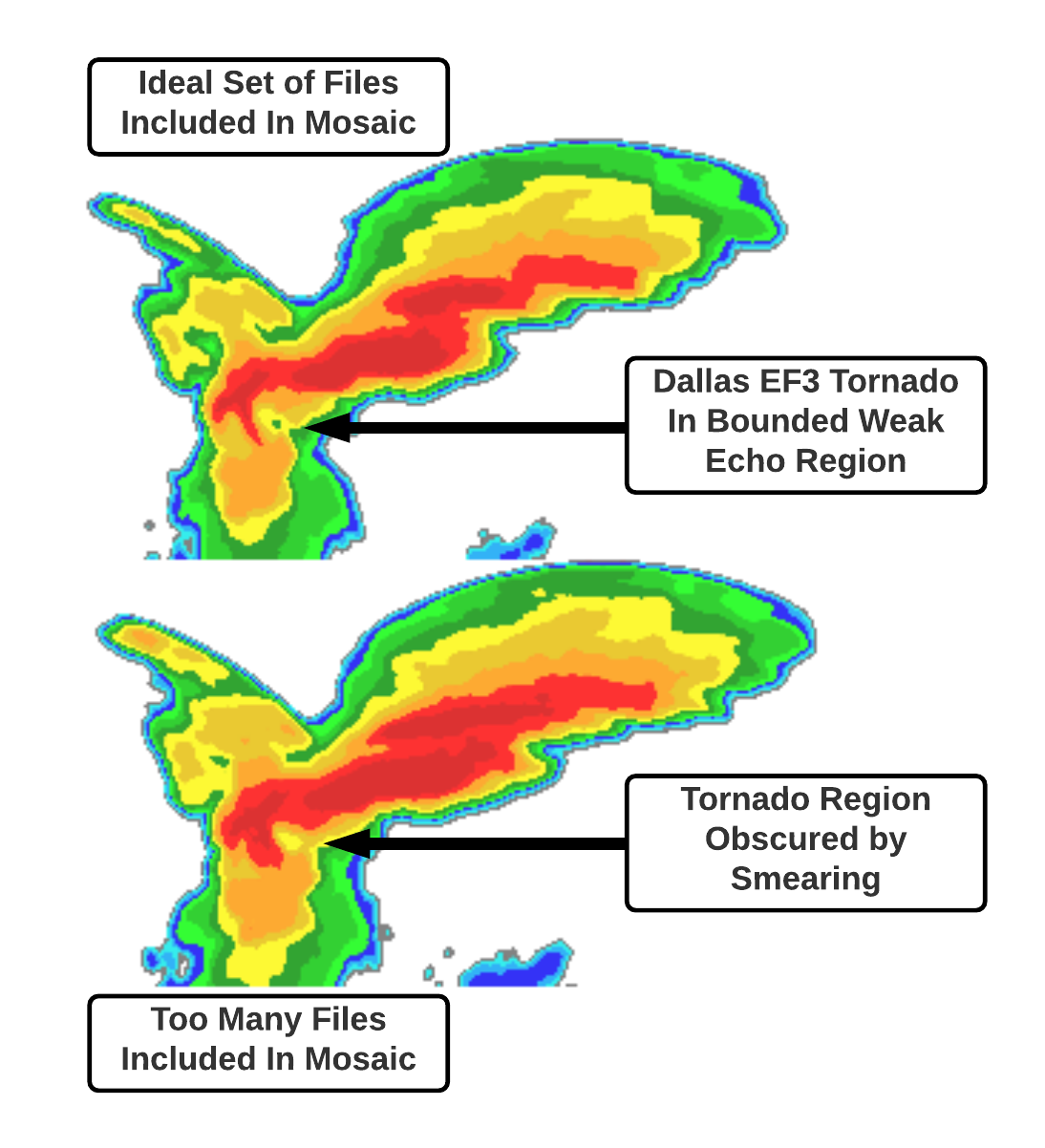}
    \caption{Merged radar data from the same sever weather event, in the case of ideal data ingest (top) and the ingest of too many files (bottom).}
    \label{fig:merged_hook}
\end{figure}

Figure \ref{fig:merged_hook} shows the effect of this change. With the current time based data ingest process, it is difficult to locate the tornado that is obscured by too many files included in the mosaic. However, with the NDN based dataflow providing the ideal number of files, it is easier to see the tornado in the rendering - this could impact the decision of a weather forecaster issuing a tornado warning or not.

%


\section{Conclusion and Future Work}
\label{sec:conclusion}

In this paper, we presented the design of a weather sensing framework over NDN. Our design conceptualizes the generation of weather sensing data from radars in rounds and features a round-based naming scheme for the retrieval of the generated data. In addition, our design utilizes a piggybacking scheme to communicate data generation changes from radars to clients. Our evaluation results show that the data-centric communication model of NDN enables the uncovering of hazardous weather events that would be hard to detect with the current CASA dataflow model. In the future, we plan to develop a prototype of this design and deploy it in the real-world CASA system. We also plan to explore the performance of our prototype on lossy links with varying delay.

\section*{Acknowledgments}

This work was funded by the National Science Foundation
(NSF) grants OAC-2019163, OAC-2126148, OAC-2019012, OAC-2018074, CNS-2104700, CNS-2016714, and CBET-2124918, the National Institutes of Health through award NIGMS/P20GM109090, the University of Nebraska Collaboration Initiative, and the Nebraska Tobacco Settlement Biomedical Research Development Funds. 
\balance
\bibliographystyle{unsrt}
\bibliography{bib,casa}

\end{document}